\documentclass[apj]{emulateapj}
\usepackage{amssymb,amsmath}
\usepackage{epsfig}

\slugcomment{Accepted to ApJ Letters}
\shorttitle{A NEW ULX IN NGC~891}
\shortauthors{HODGES-KLUCK ET AL.}

\begin{document}

\title{A New Ultraluminous X-ray Source in the Nearby Edge-on Spiral NGC~891}

\author{Edmund~J.~Hodges-Kluck$^{1}$, Joel~N.~Bregman$^{1}$,
  Jon~M.~Miller$^{1}$ \& Eric~Pellegrini$^{1}$}

\altaffiltext{1}{Department of Astronomy, University of Michigan, Ann
  Arbor, MI 48109}
\email{hodgeskl@umich.edu}

\begin{abstract}
We report the discovery of a new candidate ultraluminous X-ray source (ULX) in
the nearby edge-on spiral galaxy NGC~891.  The source,
which has an absorbed flux of $F_x \sim 1\times 10^{-12}$~erg~s~cm$^{-2}$
(corresponding to a $L_x \gtrsim 10^{40}$~erg~s$^{-1}$ at 9~Mpc), must
have begun its outburst in the past 5~years as it is not detected in
prior X-ray observations between 1986 and 2006.   We try empirical
fits to the \textit{XMM-Newton} spectrum, finding that the spectrum is fit very well
as emission from a hot disk, a cool irradiated disk, or blurred
reflection from the innermost region of the disk.  The simplest
physically motivated model with an excellent fit is a hot disk
around a stellar-mass black hole (a super-Eddington outburst), but equally good
fits are found for each model.  We
suggest several follow-up experiments that could falsify these models. 
\end{abstract}


\section{Introduction}
Ultraluminous X-ray sources (ULXs) are non-nuclear X-ray sources with
luminosities $L_X \gtrsim 10^{39}$~erg~s$^{-1}$
\citep{fabbiano89}.  These sources are interesting because their luminosities
exceed the Eddington limit for a $10 M_{\odot}$ black hole, suggesting that
they are either ``intermediate mass'' black holes (IMBHs) of 
$M_{\text{BH}} \sim 10^2 - 10^4 M_{\odot}$ \citep{colbert99} or 
stellar-mass black holes seen during a special time 
\citep[super-Eddington accretion; see, e.g.][hereafter GRD09]{gladstone09} or at a
special angle \citep[i.e., non-isotropic $L_X$;][]{king01}.  In any
case, they are important sources for studying black hole physics.  For a recent
review, see \citet{feng11}.  

In this Letter, we report the appearance of a new ULX candidate in the nearby
isolated, edge-on spiral galaxy NGC~891.  The galaxy is thought to be a
Milky Way analog in luminosity \citep{devaucouleurs91} and color \citep{vanderkruit81},
and, like the Galaxy, it is a barred spiral \citep[e.g.][]{garcia-burillo95}.
In Section~2, we describe the \textit{XMM-Newton} detection and other
observations, in Section~3 we attempt empirical fits to the X-ray
spectrum, and in Section~4 we discuss our results in context.  

Throughout this Letter, we use a Galactic column of $N_H =
6.5\times10^{20}$~cm$^{-2}$ \citep{kalberla05} as well as the
\textit{Wilkinson Microwave Anisotropy Probe} cosmology 
\citep[$H_0 = 71$~km~s$^{-1}$~Mpc$^{-1}$, $\Omega_{\Lambda} = 0.73$, and 
$\Omega_m = 0.27$;][]{spergel07}.  NGC~891 is about 9-10~Mpc away
\citep{temple05,tully09}, and we adopt $d\approx 9$~Mpc.

\section{A New Candidate ULX}


\begin{figure*}[t]
\begin{center}
\includegraphics[width=0.63\linewidth]{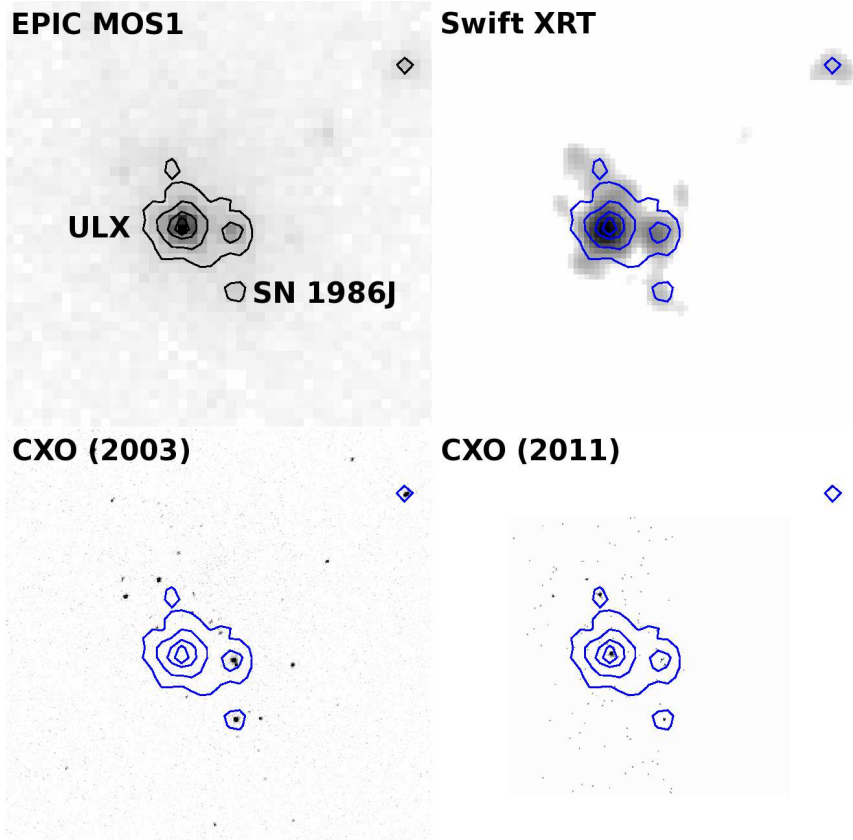}
\caption{$0.3-10$~keV X-ray maps of the field.  
\textit{Top Left}: EPIC-MOS1 image with bright contours of the same (overlaid on other images).  
\textit{Top Right}: Combined \textit{Swift} XRT image from our monitoring campaign (21~ks),
smoothed by three pixels.  \textit{Bottom Left}:
2003 \textit{Chandra} image (120~ks).  \textit{Bottom Right}: 2011 
\textit{Chandra} snapshot (2~ks).  Note the additional new source to the north also seen by
\textit{XMM-Newton} has a $0.3-10$~keV $F_X \sim 2\times 10^{-13}$~erg~s$^{-1}$~cm$^{-2}$. }
\label{xray_map}
\end{center}
\end{figure*}

A new bright source was discovered in a 133~ks \textit{XMM-Newton} exposure of NGC~891 (obsID
0670950101) with an aperture-corrected $0.3-10$~keV EPIC-pn flux of $F_X = 1.6\times
10^{-12}$~erg~s$^{-1}$~cm$^{-2}$.   
An inspection of prior X-ray observations between 1994 and 2006,
including high resolution \textit{Chandra} and ROSAT observations,
demonstrates that this source had not yet been detected at sensitivities
down to $F_X \lesssim 10^{-15}$~erg~s$^{-1}$~cm$^{-2}$ in the $0.3-10$~keV band.  
A 2~ks \textit{Chandra}
observation (obsID 14376) indicates source coordinates (J2000) 
02:22:33.45$\pm$0.03 +42:20:26.8$\pm$0.5 after processing with the 
sub-pixel EDSER algorithm \citep{li04}.

\begin{figure}[t]
\begin{center}
\includegraphics[width=0.55\linewidth,angle=90]{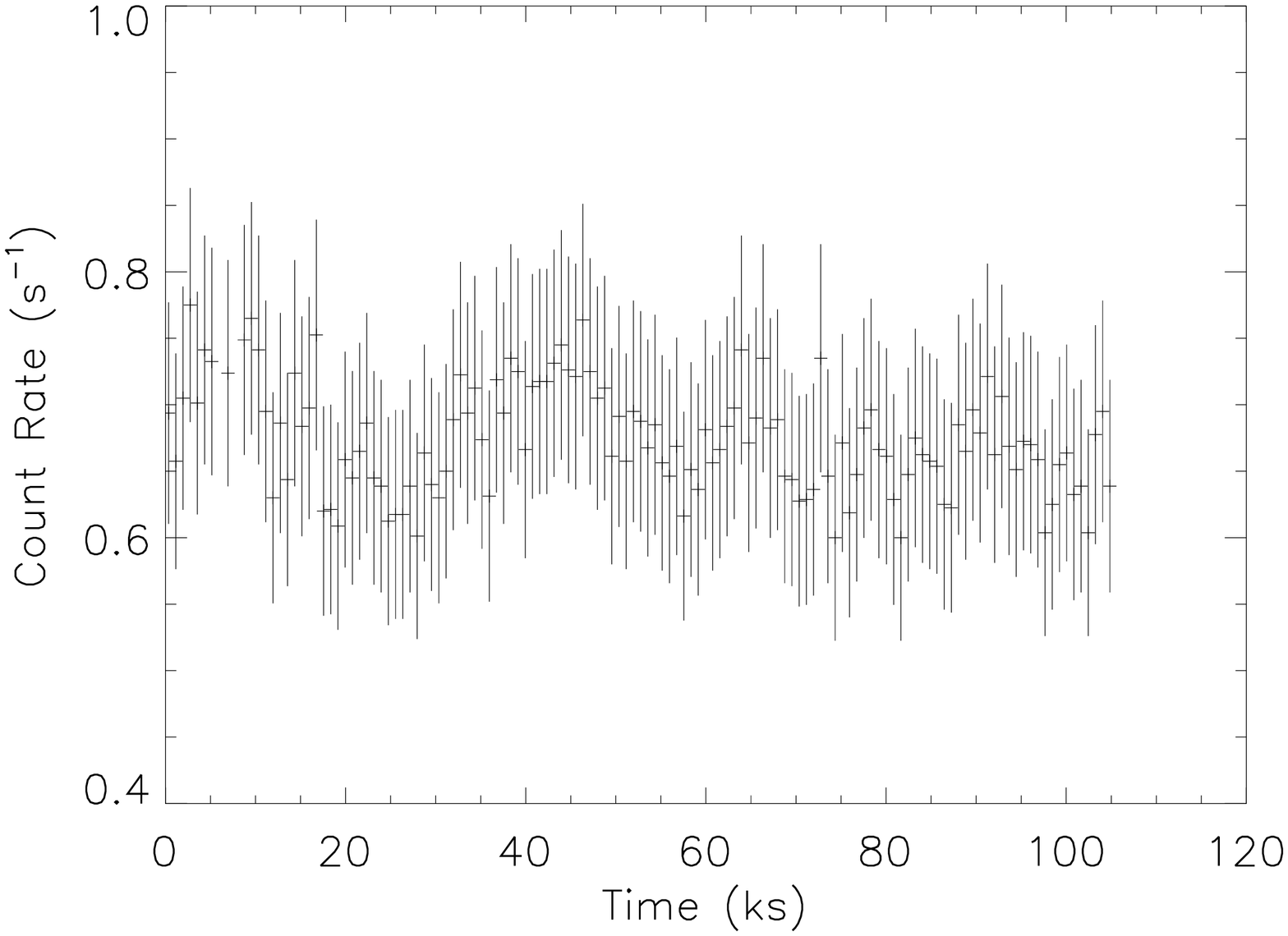}
\includegraphics[width=0.55\linewidth,angle=90]{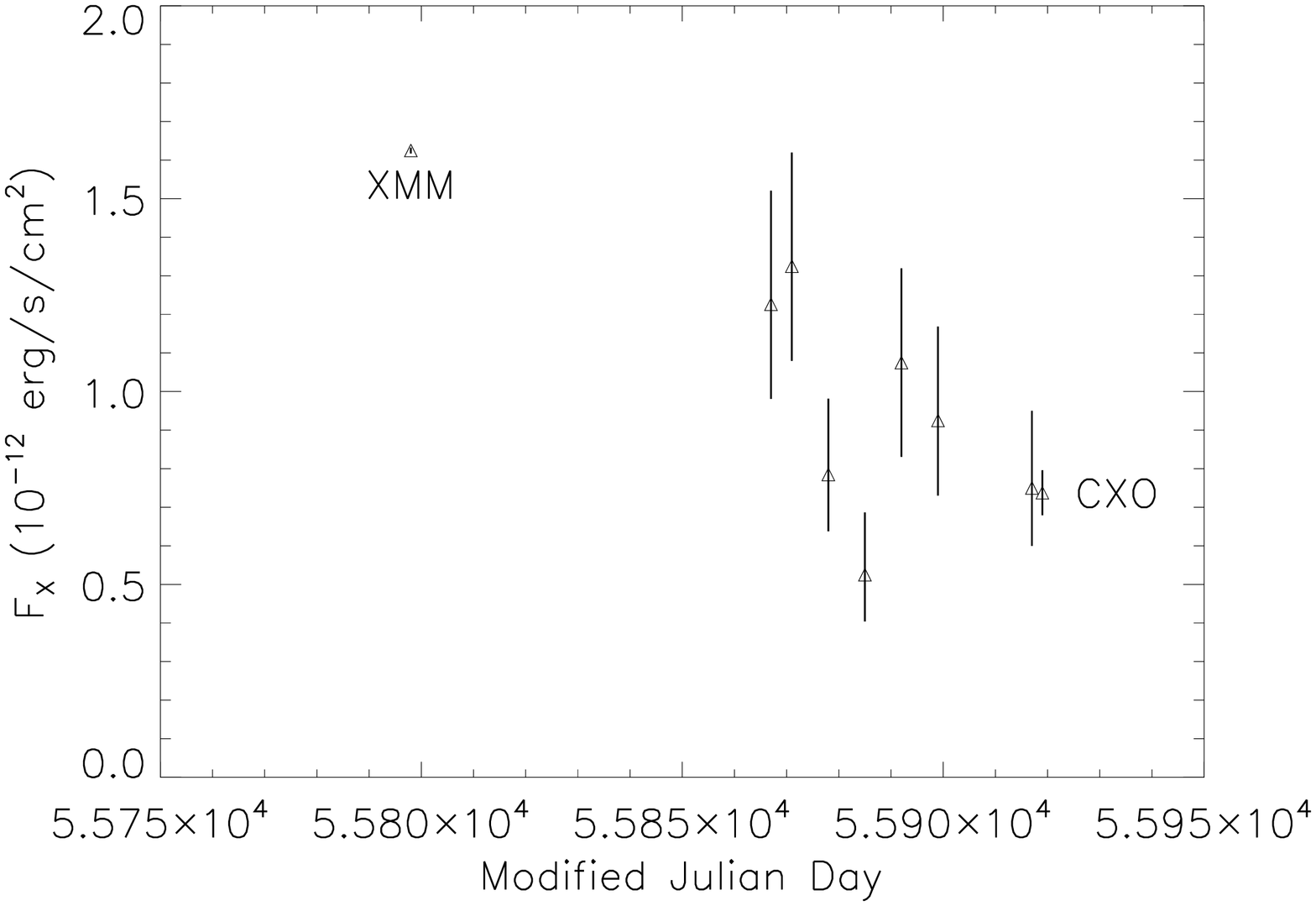}
\caption{\textit{Left}: Combined synchronized EPIC light curve binned
  to 800~s in GTIs.  \textit{Right}: \textit{XMM-Newton}, \textit{Swift}, and
  \textit{Chandra} fluxes measured since 08/2011 with 90\% error bars.}
\label{flux}
\vspace{-0.5cm}
\end{center}
\end{figure}

Flux monitoring with the \textit{Swift} XRT (target ID 35869) indicates an
overall decay with some hint of variability (Figure~\ref{flux}), but further
monitoring is necessary to establish a decay.  The \textit{Swift} values include
a minor correction for contamination by a nearby X-ray binary with a separation
of about the \textit{Swift} point-spread function, but Poisson noise is the
dominant source of uncertainty. 

Follow-up optical observations, including concurrent \textit{XMM-Newton}
optical monitor and \textit{Swift} UVOT exposures as well as a snapshot
with the Mayall~4m telescope at Kitt Peak National Observatory, detect no
optical counterpart to a limiting magnitude $V \sim 20$~mag.  NGC~891 is 
regularly monitored for supernova candidates, with no detections in the past
decade and regular exposures in the past six months (A.~Filippenko, private
communication).  Likewise, a follow-up 5~GHz 
observation with the EVLA at a sensitivity of 15~$\mu$Jy~bm$^{-1}$ (for a
beamsize $\theta \sim 10$~arcsec) found no counterpart.  Diffuse emission
from NGC~891 with a flux of $F_{\nu} \approx 80$~$\mu$Jy~bm$^{-1}$ is detected 
at this position, meaning any future attempt requires higher resolution. 

Spectral fits (Section~3) indicate an absorbing column several times higher than the
Galactic value, suggesting an extragalactic origin.  The source is coincident with
the disk of NGC~891, but it may be a background quasar or blazar.  Such a source
detected in the X-rays ought to be associated with a bright optical/UV counterpart 
with a flux density within an order of magnitude of the X-rays \citep[see, e.g.][]{shang11},
and we would expect rapid variability in the \textit{Swift} monitoring for
a blazar.  The absence of these signatures suggests a physical association with NGC~891,
but this is not yet certain. 


If the source is in NGC~891, the $0.3-10$~keV fluxes indicate (absorbed) luminosities
between $5-15\times 10^{39}$~erg~s$^{-1}$, suggesting a ULX or supernova.  Since we
have seen no radio counterpart, a normal Type~Ia or II supernova is disfavored.
Although we might also expect a radio counterpart to a ULX, the radio luminosities
expected are substantially smaller: using the ``fundamental plane'' expression for
radio luminosity as a function of black hole mass and X-ray luminosity in \citet{merloni03},
we expect $L_{\text{radio}} < 10^{34}$~erg~s$^{-1}$ for a $10^{4} M_{\odot}$ black hole.
Assuming synchrotron emission near 1~GHz with a spectral index $\alpha = 0.7$, the flux
density from such a source is at most $\sim$50~$\mu$Jy, already below the diffuse radio
halo in the 10~arcsec EVLA beam.  The flux would be even less for smaller black holes.
The radio non-detection is therefore inconsistent with normal supernovae and consistent
with a black hole.  Therefore, the source appears to be a new ULX.  Hereafter, we refer
to it as NGC~891 ULX1.  The most recent
X-ray observation prior to the \textit{XMM-Newton} exposure places an upper bound of
5~years on the outburst. 

\begin{figure}[t]
\begin{center}
\includegraphics[width=0.95\linewidth]{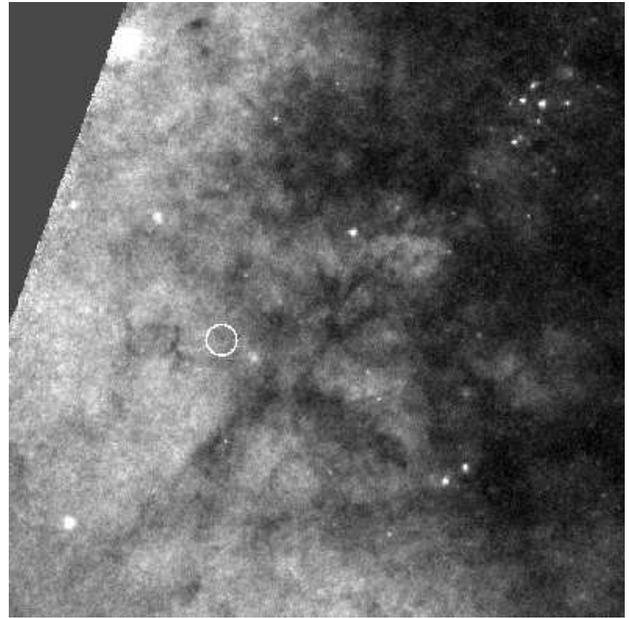}
\caption{2004-01-17 \textit{HST} ACS HRC image of NGC~891 in the F555W
  filter ($\sim$5400\AA).  Lighter color is positive emission.  The
  0.5$^{\prime\prime}$ ULX position error circle is overplotted,
  corresponding to 25~pc across.}
\label{hst}
\end{center}
\end{figure}

Archival \textit{Hubble Space Telescope} ACS HRC and WFC images from 2004 
(obsID 9414) reveal a possibly nearby source within an arcsecond of the
X-ray position, but not directly associated (Figure~\ref{hst}).  
Based on centroids of bright sources in the field, the 
\textit{HST} astrometry is uncertain to within 0.5~arcsec. 
The source is faint with a FWHM diameter of $\sim$0.23~arcsec (the PSF is
about 0.1~arcsec at FWHM), and may be a star cluster.  In the F555W filter,
it has a ST magnitude of $\sim 24.3$~mag ($M_{\text{F555W}} \sim -5.7$~mag,
corrected for Galactic $A_V \sim 0.22$~mag), 
and in the F814W filter it is greater than $23.5$~mag ($M_{\text{F814W}} > -6.4$~mag,
corrected for Galactic $A_I \sim 0.12$~mag).  This source is too dim to be seen in
the optical data described above.  If it is a cluster, the ULX may be physically
related \citep{zezas02}. 

\section{Spectral Fits}

We extracted spectra in the $0.3-10$~keV band from the EPIC-pn and MOS 
cameras aboard \textit{XMM-Newton} using standard Scientific Analysis System
recipes.  We excluded periods of background flaring during the first 10~ks and
last 30~ks, leaving 92~ks of good time intervals.  Because the field is crowded,
we extracted spectra only from within a aperture with $r = 25$~arcsec (corresponding
to about 80\% encircled energy at 1.5~keV), but reported fluxes are aperture
corrected.  The pn count rate in this aperture is 
$\sim$0.45~cts~s$^{-1}$, and the task \textit{epatplot} shows a negligible pile-up fraction.  
The MOS spectra are likewise not piled up.  

A two-sided Kolmogorov-Smirnov test finds that the light curve (Figure~\ref{flux}) is 
incompatible with a constant flux ($P < 10^{-6}$), but variability is weak, with
a fractional r.m.s. in 100~s bins of only 13\% and no evidence for quasi-periodic
oscillation in the combined, synchronized light curve.  

The spectrum is a featureless
continuum aside from the effects of the absorbing column.  In this section, we
describe empirical model fits using XSPEC v12.7.0 \citep{arnaud96}.  Each
spectrum is binned by 20~counts and we use the $\chi^2$ goodness-of-fit 
statistic.  An acceptable model is one that is not rejected at 95\% probability
by this metric.  In comparing models that fail the regularity conditions for the
\textit{F}-test \citep{protassov02}, we calibrate the \textit{F}-test in a method similar to
the ``parametric bootstrapping'' described in \citet{protassov02} by using
the best-fit model (the null model) to generate a large number (100--1000) of
fake spectra with the XSPEC `fakeit' tool for the same exposure time.  We
fit each simulated spectrum with the null model and the test model to obtain
a distribution of $\delta \chi^2$, which we compare to the $\delta \chi^2$
value for the data.  All of our spectral fits incorporate a photoelectric
absorption component with $N_H$ frozen to the Galactic value, which is not
listed with our model parameters (Table~\ref{spectralfits}), as well as an
absorption component that is free to vary.  We use {\sc TBabs} with the
\citet{wilms00} abundances.  


\begin{figure*}[t]
\begin{center}
\includegraphics[width=0.65\linewidth]{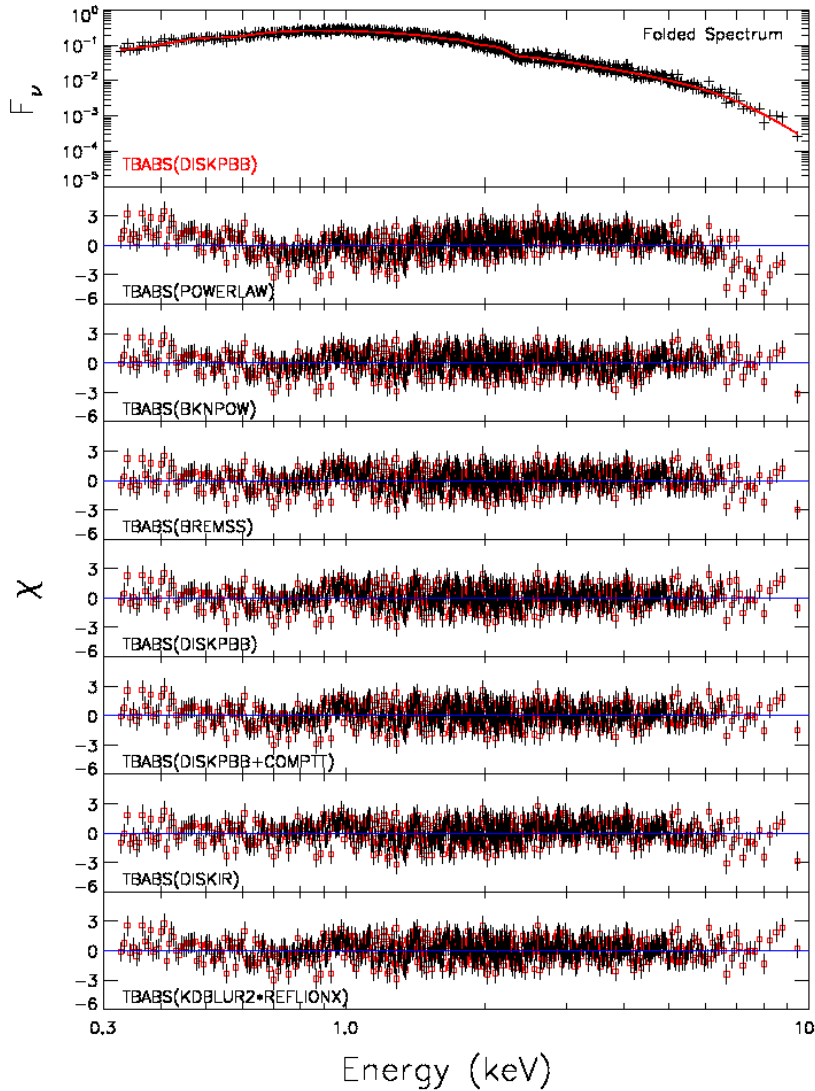}
\caption{The $0.3-10$~keV EPIC-pn spectrum (binned to 25 counts) with the best-fit
``hot disk'' model overplotted.  Below are residuals for each model.  Note the rightmost
residual for the power-law model is at $-10$.}
\label{EPICspectrum}
\end{center}
\end{figure*}

The simplest models that have been applied to ULX spectra are absorbed 
power-laws and thermal bremsstrahlung.  Although formally a poor fit ($\chi^2$/d.o.f$=1860.2/1171$), the single
power-law model ({\sc TBabs*powerlaw}) with $\Gamma = 2.24^{+0.01}_{-0.02}$ is instructive
in that its residuals show the high energy ``curvature'' often associated
with ULXs \citep{roberts06} and excess emission below 1~keV 
(Figure~\ref{EPICspectrum}).  Since Galactic black hole binaries do not
exhibit such curvature, ULXs are not simply scaled-up versions of ordinary
stellar-mass black holes.  A broken power-law ({\sc TBabs*bknpower}) with a
break energy near 4~keV is an adequate fit (Table~\ref{spectralfits}),
but it does not completely remove the excess below 0.4~keV (Figure~\ref{EPICspectrum}).
Likewise, a bremsstrahlung model ({\sc TBabs*bremss}) is a good fit
with $kT \approx 3.4$~keV but retains this excess.  
These fits can be improved by adding a cool thermal ({\sc apec}) model with fixed 
$kT = 0.1$~keV; $\delta \chi^2 = 35$ in the broken power-law model for one fewer degree of freedom indicates
an improvement at over 95\% significance.  In fact,
the excess below 0.4~keV is present in all our fits (Figure~\ref{EPICspectrum}),
but the pn and MOS disagree at these energies \citep[see examples in][]{stuhlinger06}
so we cannot conclude that the thermal component is physical and do not include it
in our reported fits.  We also exclude MOS data below 0.4~keV.  However, including
the thermal component does not cause large shifts in model parameters in 
Table~\ref{spectralfits}.


We attempted some of the disk models recently tried on high quality \textit{XMM-Newton} spectra
in GRD09 and \citet[][whose source NGC~4715 ULX1 has a qualitatively similar spectrum]{walton11a}.
These include a multi-colored disk (MCD)
blackbody model ({\sc TBabs*diskpbb}), a MCD model with Comptonization
({\sc TBabs*(diskpbb+comptt)}), a disk irradiated by Comptonized photons
({\sc TBabs*diskir}), and blurred reflection ({\sc TBabs*kdblur2*reflionx}).
In the Comptonization models, the seed photons are assumed to originate at
the inner edge of the disk. 

For the pure MCD model we use a ``$p$-free'' model in which $T(r) \propto r^{-p}$,
where $p$ is a free parameter \citep[in the normal MCD model {\sc diskbb}, $p$
is frozen at $0.5$;][]{mitsuda84}.  The best fit has an inner disk temperature 
$T_{\text{in}} \approx 1.6$~keV and $p \approx 0.54$, indicating a hot, ``slim'' disk.  
The temperature can be lowered to $T_{\text{in}} \sim 1$\,keV by adding a 
Comptonization component, but based on simulated spectra the additional component does
not significantly improve the fit.  Because the
spectrum remains disk-dominated in either case, we find a degeneracy between
a hot, optically thick corona or a cool, optically thin one.  GRD09 find that
a cool, thick corona is preferred, but we find no global minimum and poor
constraints due to the weakness of the component. 

We obtain a good fit for a cool disk with a dominant Comptonization component
when the disk structure is modified by the absorbing Comptonized photons
({\sc diskir}).  In this model, the inner disk is irradiated by Comptonized photons,
many of which are ``reflected''.  Some fraction is absorbed and re-emitted with
a quasi-thermal spectrum, thereby modifying the disk emission \citep{gierlinski08}.
Significant modification only occurs when the Comptonized component is much 
brighter than the disk emission, naturally forcing fits with cool, relatively
dim disks.  Thus, this model is expected to be important in the low/hard state---a
very different situation from that implied by the MCD model.  In order to fit
this spectrum,the Compton hump must occur near 2~keV, implying a cool corona
($kT_e \sim 1.4$~keV) and a cooler disk ($T_{\text{in}} \sim 0.4$~keV).  The
intrinsic absorption is also substantially lower than in other models.

It is also possible that much of the emission is Comptonized photons
``reflected'' off the inner edge of the disk with substantial line ``blurring''
due to relativistic effects near the black hole.  Like \citet{walton11a},
we find a significant amount of blurring is required such that the
disk must reach to 1.26$R_G$ with an emissivity index $q_{\text{in}} > 7$,
meaning that the emission is dominated by light from the inner disk.  

\section{Discussion}

Despite the high quality of the X-ray spectrum, several very different models
produce excellent fits.  Thus, additional data at other wavebands or a timing
study in the X-rays are required to discriminate between them.  We briefly
discuss the implications and predictions of each model.

\paragraph{\textit{A Hot Disk}}
The hot disk model implies a black hole mass of less than $40 M_{\odot}$ 
\citep[e.g.][]{soria08}, hence suggesting super-Eddington accretion.  Similar hot-disk
ULXs are explained as stellar-mass black holes by \citet{winter06}, and the recent appearance of
the source makes sense if, as expected, super-Eddington accretion is a transient
phenomenon.  The \textit{Chandra} non-detection places a prior limit of $L_X < 10^{37}$~erg~s$^{-1}$,
which is consistent with a low/hard state in a stellar-mass black hole but quiescence in an IMBH.
The excellent fit with a small number of parameters is also a point in favor
of the pure MCD model, and a disk-dominated spectrum is consistent with an
extreme version of the high/soft state seen in Galactic black holes
\citep[][GRD09]{mcclintock06}. 


\begin{deluxetable}{llll}
\tablenum{1} 
\tabletypesize{\scriptsize} 
\tablecaption{Spectral Models} 
\tablewidth{0pt} 
\tablehead{ 
\colhead{Component} & \colhead{Parameter} & \colhead{Units} & \colhead{Value} 
}
\startdata
\multicolumn{4}{c}{\sc TBabs*bremss}\\
\cline{1-4}\\
TBABS     & $N_H$         & $10^{21}$ cm$^{-2}$   & $1.3\pm0.1$ \\
BREMSS    & $kT$          & keV                   & $3.4\pm0.1$ \\
$\chi^2$ (d.o.f.) & &                       & 1189.3 (1171)\\
\cline{1-4}\\
\multicolumn{4}{c}{\sc TBabs*bknpower}\\
\cline{1-4}\\
TBABS     & $N_H$         & $10^{21}$ cm$^{-2}$   & $1.9\pm0.1$ \\
BKNPOWER  & $\Gamma_1$    &                       & $1.91\pm0.04$ \\
          & $\Gamma_2$    &                       & $3.2\pm0.2$ \\
          & BreakE        & keV                   & $3.7^{+0.2}_{-0.3}$ \\
$\chi^2$ (d.o.f.) & &                       & 1217.0 (1169)\\
\cline{1-4}\\
\multicolumn{4}{c}{{\sc TBabs*diskpbb} (Hot Disk)}\\
\cline{1-4}\\
TBABS     & $N_H$           & $10^{21}$ cm$^{-2}$   & $1.5\pm0.1$ \\
DISKPBB   & $T_{\text{in}}$ & keV                   & $1.62^{+0.05}_{-0.06}$ \\
          & $p$             &                       & $0.54\pm0.01$ \\
$\chi^2$ (d.o.f.) &   &                       & 1169.1 (1170) \\
\cline{1-4}\\
\multicolumn{4}{c}{{\sc TBabs*(diskpbb+comptt)} (Hot Disk)}\\
\cline{1-4}\\
TBABS     & $N_H$           & $10^{21}$ cm$^{-2}$   & $1.5\pm0.1$ \\
DISKPBB   & $T_{\text{in}}$ & keV                   & $1.1\pm0.1$ \\
          & $p$             &                       & 0.54 (f)\\
COMPTT    & $kT_e$          & keV                   & $44$\tablenotemark{a} \\
          & $\tau_p$        &                       & $0.01$\tablenotemark{a}\tablenotemark{b}\\
\textit{or}& $kT_e$         & keV                   & 2.0\tablenotemark{a}\tablenotemark{b}\\
          & $\tau_p$        &                       & 5.4\tablenotemark{a}\\
$\chi^2$ (d.o.f.) &   &                       & 1166.6 (1168)\\
\cline{1-4}\\
\multicolumn{4}{c}{{\sc TBabs*diskir} (Cool Disk)}\\
\cline{1-4}\\
TBABS     & $N_H$           & $10^{21}$ cm$^{-2}$   & $0.8^{+0.5}_{-0.3}$ \\
DISKIR    & $T_{\text{in}}$ & keV                   & $0.38_{-0.03}^{+0.04}$ \\
          & $\Gamma$        &                       & $1.93_{-0.02}^{+0.05}$\tablenotemark{b} \\
          & $L_c/L_d$       &                       & $7^{+4}_{-2}$ \\
		  & $kT_e$          & keV                   & $1.4\pm0.3$ \\
          & $f_{\text{in}}$ &                       & 0.1 (f)\\
          & $f_{\text{out}}$&                       & 0.04$^{+0.06}$\tablenotemark{b}\\
          & $r_{\text{irr}}$& $r_{\text{in}}$       & $6^{+3}_{-2}$\\
          & $r_{\text{out}}$& $r_{\text{in}}$       & $10^5$ (f)\\
$\chi^2$ (d.o.f.) &   &                       & 1160.8 (1166)\\
\cline{1-4}\\
\multicolumn{4}{c}{{\sc TBabs*kdblur2*reflionX} (Blurred Reflection)}\\
\cline{1-4}\\
TBABS     & $N_H$           & $10^{21}$ cm$^{-2}$   & $2.1\pm0.2$ \\
KDBLUR2   & $R_{\text{in}}$ & $R_G$                 & $1.26^{+0.08}$\tablenotemark{b}\\
          & $R_{\text{out}}$& $R_G$                 & 400 (f)\\
          & $i$             & deg                   & $30\pm11$ \\
          & $q_{\text{in}}$ &                       & $8^{+2}_{-1}$ \\
          & $q_{\text{out}}$&                       & 3.0 (f)\\
          & $R_{\text{break}}$ & $R_G$              & 20.0 (f)\\
REFLIONX  & $\Gamma$        &                       & $1.73\pm0.06$ \\
          & $\xi$           & erg cm s$^{-1}$       & $>6000$\tablenotemark{b} \\
          & $A_{\text{Fe}}$ &                       & $8\pm3$ \\
$\chi^2$ (d.o.f.)   &  &                      & 1242.34 (1167)\\
\enddata
\tablenotetext{a}{These parameters are not well constrained and we do
  not quote errors.  See text.}
\tablenotetext{b}{These parameters are near the boundary of parameter space.}
\tablecomments{\label{spectralfits} All fits incorporate a 
  {\sc TBabs} component frozen at the Galactic $N_H = 6.5\times
  10^{20}$~cm$^{-2}$.  Additional
  absorption is listed here in units of $10^{21}$~cm$^{-2}$.  Errors
  are quoted at the 90\% confidence interval based on the XSPEC task \textit{steppar}. }
\end{deluxetable}

The unabsorbed model luminosity is $L_X = 2.1\pm0.3 \times 10^{40}$~erg~s$^{-1}$, 
significantly higher than other disk-dominated
ULXs \citep{swartz04,swartz03,roberts02} and modestly higher than the brightest hot-disk
sources in \citet{winter06}.  This luminosity exceeds the Eddington limit by 2.5~times for
a $100 M_{\odot}$ black hole.  Hence, it does not easily 
fit into the ``sequence'' proposed by GRD09, in which disk-dominated sources
are sub-Eddington accretors on the extreme tail of the stellar-mass black hole mass distribution.
As the source maintains the same hardness ratio seen in the \textit{XMM-Newton} spectra during
\textit{Swift} monitoring, there may be substantial scatter in $L_X - T_{\text{in}}$ plots
\citep[cf.][]{miller04}.  If the source is thermally dominated (the simplest model with an
excellent fit), it would join only a few other ULXs seen in this state
\citep{feng10,jin10,servillat11}.  

In the case of super-Eddington accretion we expect to see powerful
outflows.  Since the outburst began at most 5 years ago, these outflows
would be within $\sim$1~pc of the source.  We might therefore expect to 
see absorption signatures in a high quality RGS spectrum, although the source
geometry is unknown.  Detection of such an outflow would place a strong
constraint on the mechanical luminosity, an important parameter in models
of super-Eddington accretion \citep[e.g.][]{begelman02,dotan11}.  If 
super-Eddington accretion in this system is episodic, broad or multiple
absorption features would be expected.  

\paragraph{\textit{A Cool Disk}}
If the hot disk model is similar to an extreme version of the high state
seen in Galactic black holes, a cool disk may correspond roughly
to the low/hard state \citep{gierlinski08}.  In the irradiated disk model, the spectrum is
dominated by a cool, Comptonized component with the subjacent disk
component dominant at low energies.  This model presumes strong reflection,
but the absence of a strong Fe~K$\alpha$ line is consistent with the low
corona temperature ($kT_e \approx 1.4$~keV).  The emission comes almost
exclusively from the very inner region of the disk, which itself is
cool ($T_{\text{in}} \lesssim 0.4$~keV).  If this disk extends to the
innermost stable circular orbit, its temperature would indicate
an IMBH (and we would expect the luminosity to vary as $L_X \propto T_{\text{in}}$ in
future observations).  
However, the disk may be truncated at large radii outside of a
large corona or outflow \citep[further discussed in][]{feng11}, in which case the
disk temperature cannot be used to infer a mass.  

There are a few objections to this model.  First, a cool disk component is
only a good fit when irradiated, otherwise a hot disk is required (even the ``cooler'' hot
disk has $T_{\text{in}} \sim 1$\,keV).  
If the disk were intrinsically cool, we would expect a non-irradiated
cool disk to at least be competitive with the hot disk.  Second, if
the black hole is in the low/hard state, the prior X-ray non-detections
suggest it was previously quiescent; even the naked disk in this model
should have been detected.  

Fortunately, this model can be falsified in a few ways.  The intrinsic
absorption in this fit is small compared to the others, so if an optical
counterpart could be established and its intrinsic color determined, a
column could be measured.  We would also expect to see variability like
that in Galactic low/hard states \citep[and potentially quasi-periodic
oscillations as in][]{strohmayer09}, so long-term X-ray monitoring can
potentially discriminate between a high and low state. 

\paragraph{\textit{Blurred Reflection}}
In contrast to the other models, the blurred reflection model fits the
spectrum assuming it is dominated by reflection of Comptonized photons
near the inner edge of the disk, with relativistic effects smearing
out the strong emission lines.  This model predicts a Compton hump in
the 10--30~keV range, so observations above 10~keV could easily
distinguish between this model and the others.  Whereas many ULXs are in 
galaxies with nuclear sources of comparable or greater brightness, NGC~891 ULX1
is presently the brightest X-ray source in the galaxy.  As in the irradiated
disk model, blurred reflection suggests the low/hard state.  The
caveats of the physical interpretation here are discussed in detail in
\citet{walton11a}, but we note here that this model requires suppression of
the disk emission. 

\ \\ 
Like other ULX spectra, the mostly featureless continuum of NGC~891 ULX1
admits several different scenarios, suggesting observations in other
wavebands or in the time domain are necessary to untangle the origin of the X-rays.
Given that the outburst is recent and preceded by at least a 20~year lull, a
search for an ionized nebula like those seen around other nearby, bright ULXs
\citep{pakull02} is of interest.  Such a nebula would point to recurrent
ultraluminous activity, and its size and power may allow measurement of the
duty cycle.

\section{Summary}

A recent \textit{XMM-Newton} observation revealed a bright new source towards
NGC~891.  At this point, the most natural explanation is that it is a source
within the galaxy, in which case its flux and the absence of detections in other
wavelengths makes a ULX the best explanation.  The \textit{XMM-Newton} spectrum
is morphologically similar to disk-dominated sources in the GRD09 ``sequence'' of
ULXs and is indeed fit well by a hot MCD disk.  However, several very different
physical scenarios produce good fits to the spectrum, including an irradiated
cool disk with strong but cool Comptonization and a blurred reflection model.
In the hot-disk model, NGC~891 ULX1 has a mass less than $40 M_{\odot}$ and therefore
accretion rates over 5~times Eddington.  If so, the recent ignition suggests a
search for outflows would be a worthwhile test of super-Eddington models.

\acknowledgments

The authors thank the referee for catching errors and making comments that significantly
improved the paper, as well as R.~Mushotzky for helpful comments and A.~Filippenko for
information regarding SNe monitoring.  
EHK gratefully acknowledges support from NASA ADAP grant \#061951.

{\it Facilities:} \facility{XMM}, \facility{EVLA}, \facility{Swift (XRT)},
\facility{CXO (ACIS)}, \facility{Mayall}, \facility{HST (ACS)}


\end{document}